\documentclass[aps,prb,preprint,superscriptaddress]{revtex4-2}
\usepackage{amsmath,amsthm,amssymb,amsfonts}
\usepackage{siunitx}
\usepackage{color}
\usepackage{hyperref}
\usepackage{diagbox}
\usepackage{graphicx} 
\usepackage{siunitx}
\usepackage{textcomp}
\usepackage{gensymb}
\hypersetup{colorlinks=true,linkcolor=blue,anchorcolor=blue,citecolor=blue}

\newcommand{\PTO}{PbTiO$_3$~}
\begin{document}

\title{\textbf{\mbox{Ultrafast Polar Topology Reversal via THz-Driven Inertial Switching}}}

\author{Denan Li}
\affiliation{Department of Physics, School of Science, Westlake University, Hangzhou 310030, China}

\author{Shi Liu}
\email{liushi@westlake.edu.cn}
\affiliation{Department of Physics, School of Science, Westlake University, Hangzhou 310030, China}
\affiliation{Institute of Natural Sciences, Westlake Institute for Advanced Study, Hangzhou 310024, China}

\begin{abstract}
Topological polar structures in ferroelectric heterostructures offer functional states beyond ordinary domains. As polarization reversal underpins ferroelectric devices, switching between degenerate topological variants, which we refer to as topology reversal, would greatly expand the ways to control polar textures. However, polar topology reversal is challenging because it requires coordinated flipping of oppositely oriented local polarizations, a process not naturally driven by a uniform electric field.
Using a multiscale modeling approach, we show that polar topology can be stabilized and reversed in a simple Pt/PbTiO$_3$/Pt nanocapacitor. Electrode-controlled boundary conditions drive tetragonal PbTiO$_3$ into competing orthorhombic- and rhombohedral-like polarization states, which form building blocks for diverse topological structures, including vortex arrays, skyrmion lattices, dipole spirals, and dipole waves. We further show that the same out-of-plane THz pulse can reversibly switch a double-$\mathbf q$ dipole wave between degenerate topological variants within a few picoseconds, while preserving its spatial modulation. This process is enabled by damping-assisted inertial switching, in which the THz pulse injects energy to drive large-angle polarization rotation across shallow barriers, while dissipation removes excess energy and traps the system in the reversed state. These results establish ferroelectric nanocapacitors as a simple platform for programmable polar topology and low-power, THz-speed control through dissipative dynamics.
\end{abstract}

\maketitle
\clearpage

%% Introduction
Ferroelectric heterostructures provide a platform for emergent topological polar structures, including flux-closure domains~\cite{Tang15p547}, vortices~\cite{Yadav16p198}, skyrmions~\cite{Das19p368}, merons~\cite{Wang20p881}, and other spatially modulated states~\cite{Lu18p177601,Gong21p5503,Hong26p1022}. 
Stabilizing these textures has generally required deliberate materials and structural design. A representative example is PbTiO$_3$/SrTiO$_3$ superlattices, where the lattice periodicity, layer thickness, and in-plane strain must be finely tuned to create the necessary energetic competition needed to enable polar topologies~\cite{damodaran17p1003,hsu19p1901014,dai23p052903}.  Topological polar textures can also emerge in capacitor geometries. 
For example, nontrivial double-$\mathbf{q}$ polar order, a polarization pattern formed by two periodic modulations along different directions, has been observed in SrRuO$_3$/PbTiO$_3$/SrRuO$_3$ capacitors~\cite{rusu22p240}. Its formation was associated with oxygen-octahedral rotations and an electric Dzyaloshinskii–Moriya-like interaction at the perovskite electrode/ferroelectric interface. Overall, the need for precisely engineered multilayers, strain states, or interface-specific couplings remains a barrier to realizing polar topologies in simple two-terminal metal/ferroelectric/metal (MFM) devices, the most common and technologically relevant geometry in electronics.

\begin{figure*}
\centering
\vspace*{-5mm}
\includegraphics[width=\textwidth]{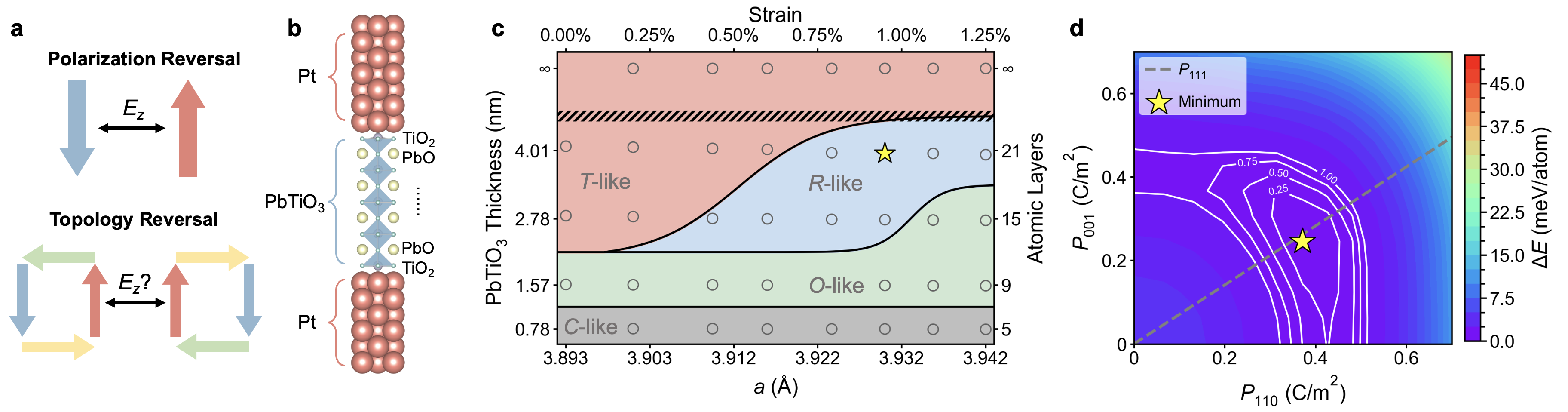}
\caption{
\textbf{Topology reversal and DFT phase diagram of Pt/PbTiO$_3$/Pt nanocapacitors.}
(a) Schematic comparison between conventional polarization reversal (upper) and topology reversal (lower). An out-of-plane field $E_z$ can naturally drive uniform polarization reversal, but whether it can induce topology reversal, which requires collective inversion of all local dipoles, is nontrivial.
(b) Atomic structure of the capacitor used in the DFT calculations, consisting of a PbTiO$_3$ film confined between Pt electrodes with TiO$_2$-terminated interfaces.
(c) Ground-state polarization phase diagram as a function of PbTiO$_3$ thickness and in-plane lattice constant. Open circles denote the calculated structures. The hatched break separates finite-thickness capacitors from the bulk limit. The strain $\eta$ is defined relative to the bulk in-plane lattice constant $a_0=3.893$~\AA.
(d) Potential energy surface as a function of the in-plane diagonal polarization $P_{110}$ and the out-of-plane polarization $P_{001}$ for a PbTiO$_3$ film containing 21 atomic layers (4.01~nm) at $\eta=0.95\%$, corresponding to the yellow star in (c). The dashed line indicates the [111] direction, and the star marks the energy minimum. The energies are normalized by the number of atoms in the \PTO layer of the nanocapacitor.
}
\label{fig:dft}
\end{figure*}

A major motivation for creating topological polar structures is their potential use as functional units in electronic devices, making deterministic electrical control essential. The ability to switch a given polar topology into its reversed counterpart is thus a key requirement. We term this process \textit{topology reversal}, in analogy to conventional ferroelectric \textit{polarization reversal}, where the polarization switches between two energetically degenerate states (Fig.~\ref{fig:dft}a, upper panel). However, polar topology reversal would require all local dipoles to reverse direction, while the real-space spatial modulation remains preserved (Fig.~\ref{fig:dft}a, lower panel).
A simple symmetry argument suggests that achieving such reversal with a spatially uniform electric field is difficult, if not impossible. Nontrivial polar textures often contain antiparallel or continuously rotating local polarizations, as in vortices. A uniform field, being a single vector, cannot simultaneously favor 180$^\circ$ switching of all local dipoles. In reciprocal space, this reflects that polar textures are finite-$\mathbf{q}$ polarization modulations, whereas a uniform electric field couples only to the $\mathbf{q}=0$ component. Consistent with this limitation, reported electrical responses of polar textures have largely been volatile, including weak-field collective oscillations of vortices and strong-field destructive transformations among labyrinth domains, skyrmion crystals, and uniformly polarized states~\cite{Li21p376,Yuan23p226801,wang25p8994,li25p1965}. These responses disappear after the field is removed and do not yield nonvolatile topology reversal. 

Here, we find that a simple two-terminal Pt/PbTiO$_3$/Pt nanocapacitor enables ultrafast control of polar topology. First-principles calculations reveal that imperfect interfacial screening, epitaxial strain, and thickness tuning can overcome the strong tetragonal preference of PbTiO$_3$, producing a softened energy landscape with unconventional non-tetragonal states, including a rhombohedral-like state. These competing polarization states provide the building blocks for polar topology: large-scale molecular dynamics (MD) simulations show that the softened landscape stabilizes diverse polar textures at finite temperature, including vortices, skyrmions, dipole spirals, and dipole waves. 
We further demonstrate reversible THz switching of a double-$\mathbf{q}$ dipole-wave topology with alternating in-plane and out-of-plane polarization components. Strikingly, an out-of-plane THz pulse reverses nearly all local dipoles by approximately 180$^\circ$ within a few picoseconds while preserving the spatial modulation. Applying the same pulse again switches the topology back, demonstrating nonvolatile topology reversal.
This counterintuitive reversal is enabled by a damping-assisted inertial switching mechanism, in which the pulse drives polarization rotation across the energy barrier while dissipation traps the system in the reversed state. These results reveal the unexpected potential of simple nanocapacitors for ultrafast control of polar topology.

%Fig 1
We begin from a simple design principle: exploit the residual depolarization field, normally detrimental in MFM nanocapacitors, to soften the ferroelectric energy landscape, and combine it with in-plane strain to stabilize new polar phases. Incomplete electrode screening can suppress ferroelectricity in the ultrathin limit~\cite{Junquera03p506}.
We instead use an intermediate-thickness regime in which the depolarization field weakens the strong tetragonal preference of \PTO without eliminating polarization. In-plane tensile strain further promotes in-plane polarization components, creating competition among multiple polarization variants~\cite{Pertsev98p1988}.
We test this principle using density functional theory (DFT) calculations on a prototypical symmetric Pt/PbTiO$_3$/Pt nanocapacitor with TiO$_2$-terminated interfaces (Fig.~\ref{fig:dft}b)~\cite{umeno09p205122,mendez12p214107,sai05p020101,kolpak06p054112,al10p155304}. Films with $N$ atomic layers are denoted as $N$L. For each thickness $t$ and strain $\eta$ relative to $a_0 = 3.893~\text{\AA}$, 
we relax the atomic coordinates and out-of-plane cell parameter from multiple initial polarization directions and identify the lowest-energy state (see Supplemental Sect.~I for computational details; the Supplemental Material includes Refs.~\cite{cp2kjcp2020,Perdew08p136406,Goedecker96p1702,Hartwigsen98p3641,Monkhorst76p5188,ferrodispcalc,yang24p214112,Batatia22p11423,Zhang20p107206,Zhang18p143001,Zeng23p054801,Shirane51p265,Gong21p5503,rusu22p240,streiffer98p2742,huang09p140101,Hu24p046802,Thompson22p108171}).

The resulting DFT ground-state phase diagram is shown in Fig.~\ref{fig:dft}c. In addition to the nonpolar state ($C$-like) and the conventional [001]-polarized state ($T$-like), we identify two additional polar states labeled by their polarization directions: [110]-polarized ($O$-like) and [111]-polarized ($R$-like). Even at zero strain, increasing thickness drives the system out of the nonpolar $C$-like ultrathin limit and first stabilizes the $O$-like state before the $T$-like state is recovered at larger thickness.
The $R$-like state emerges when appropriate tensile strain and film thickness are combined, occupying an intermediate region between the $O$-like and $T$-like states.
Although previous studies have reported that in-plane tensile strain can induce an $R$-like state in bulk PbTiO$_3$, the required strain is typically of order $\approx 2\%$~\cite{Hu24p046802}. In the nanocapacitor, the residual depolarization field softens the energy landscape, so a modest tensile strain of only $\approx 0.4\%$ is sufficient to stabilize the $R$-like state at a suitable thickness of $\approx$2.7~nm. 

Notably, the $R$-like state is both strongly polar and highly susceptible. As shown by the potential energy surface (PES) of a representative 21L \PTO film ($t=4.01$ nm and $\eta=+0.95\%$) in Fig.~\ref{fig:dft}d, the [111]-polarized ground state has a large polarization of 0.45 C/m$^2$, about $56\%$ of the bulk value. Yet the barriers separating the [001]-, [110]-, and [111]-polarized states are below 1 meV/atom, indicating an exceptionally shallow rotational energy landscape. For comparison, rhombohedral BaTiO$_3$ has a smaller polarization of 0.34 C/m$^2$, but polarization rotation toward [001] still crosses a higher barrier of $\approx 5$ meV/atom. Similarly soft landscapes are found for other representative thicknesses and strains (Fig.~S2). Thus, Pt/PbTiO$_3$/Pt nanocapacitors realize a rare combination of strong polarity and rotational softness, two properties that are often negatively correlated~\cite{Junquera23p025001}. 

\begin{figure*}
\centering
\includegraphics[width=\textwidth]{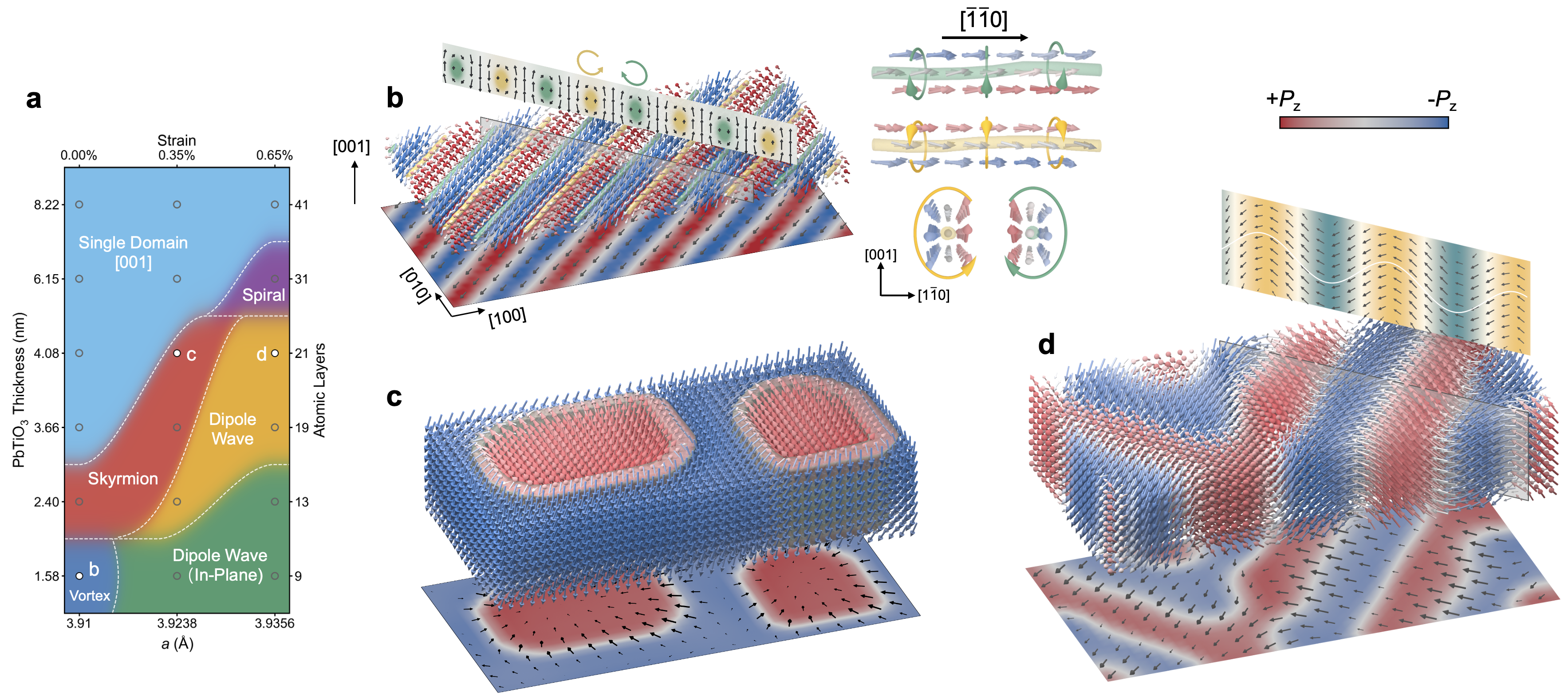}
\caption{
\textbf{Emergent polar topologies in Pt/PbTiO$_3$/Pt nanocapacitors.}
(a) Topological polar structures as a function of PbTiO$_3$ thickness and in-plane lattice clamping. The upper axis gives the epitaxial strain $\eta$, defined relative to the 100~K bulk lattice constant $a_0=3.91$~\AA. Colored regions denote distinct topologies, and letters mark the representative configurations shown in (b--d).
(b) Vortex lattice at 9L and $\eta=0\%$. Green and yellow highlight clockwise and counterclockwise vortex-core tubes, respectively. Enlarged views on the right show polarization components along the vortex-tube axis, $[\bar{1}\bar{1}0]$, while the components within the $\{\bar{1}\bar{1}0\}$ planes form the vortex lattice.
(c) Skyrmion texture at 21L and $\eta=0.35\%$. The lower projection highlights the divergent N\'eel-type polarization profile at the skyrmion boundary.
(d) Double-$\mathbf q$ dipole-wave texture at 21L and $\eta=0.65\%$. The upper and lower projections show spatial modulations of the out-of-plane and in-plane polarization components, respectively.
The simulated nanocapacitors have an in-plane supercell size of $40\times20$, with the total number of atoms ranging from 32,800 to 96,800 depending on the PbTiO$_3$ thickness.
}
\label{fig:md}
\end{figure*}

The DFT results show that film thickness and in-plane strain create a soft zero-temperature energy landscape with competing polar variants in the nanocapacitor. At finite temperature, such near-degenerate variants can readily organize into topological polar structures, which often consist of spatially rotating polarization textures. To capture them, we developed a MACE machine-learning force field trained on first-principles data for large-scale MD simulations (Supplemental Sects.~IV and V)~\cite{Batatia22p11423}. The force field underestimates the Curie temperature by approximately 100~K, a common issue associated with the underlying PBEsol exchange-correlation functional used to construct the training database~\cite{Li26p}. The resulting 100~K phase diagram, shown in Fig.~\ref{fig:md}a, reveals a rich set of equilibrium polar topologies.
Specifically, in thin films under weak tensile strain, such as 9L at $\eta=0\%$, the local unit cells adopt $R$-like states that collectively form a vortex-array lattice (Fig.~\ref{fig:md}b). This texture features periodically alternating out-of-plane polarization and a nearly homogeneous in-plane component along $[\bar{1}\bar{1}0]$. In the perpendicular $\{\bar{1}\bar{1}0\}$ plane, the dipoles form alternating clockwise and counterclockwise rotations. Increasing the thickness at weak tensile strain reduces the depolarization field and shifts the local state toward a more $T$-like character. As a result, an ordered skyrmion lattice emerges, represented by the 21L film at $\eta=+0.35\%$ (Fig.~\ref{fig:md}c). It is noted that this skyrmion lattice is an equilibrium short-circuit texture stabilized without an applied electric field, unlike previous reports~\cite{Yuan23p226801}.

Under stronger tensile strain, the thin-film limit favors an $O$-like local state and produces a single-$\mathbf q$ dipole wave dominated by in-plane polarization, as seen at 9L and $\eta=+0.65\%$ (Fig.~S7). Interestingly, increasing the thickness to 21L at the same strain restores an $R$-like local polarization again and stabilizes double-$\mathbf q$ dipole waves (Fig.~\ref{fig:md}d), in which local dipoles exhibit wave-like rotations along two distinct in-plane directions. At still larger thickness, such as 31L, the local state becomes more $T$-like, and the topology evolves into a dipole spiral with stronger out-of-plane polarization character (Fig.~S7)~\cite{Hu24p046802}.
At 300~K, a variety of topological polar structures also emerge, albeit with reduced diversity compared with those at 100~K.
The emergence of this hierarchy of polar topologies without chemical complexity, multilayer patterning, or external electric fields thus highlights nanocapacitors as a simple and tunable platform for topological polar order.

\begin{figure*}
\centering
\includegraphics[width=1\textwidth]{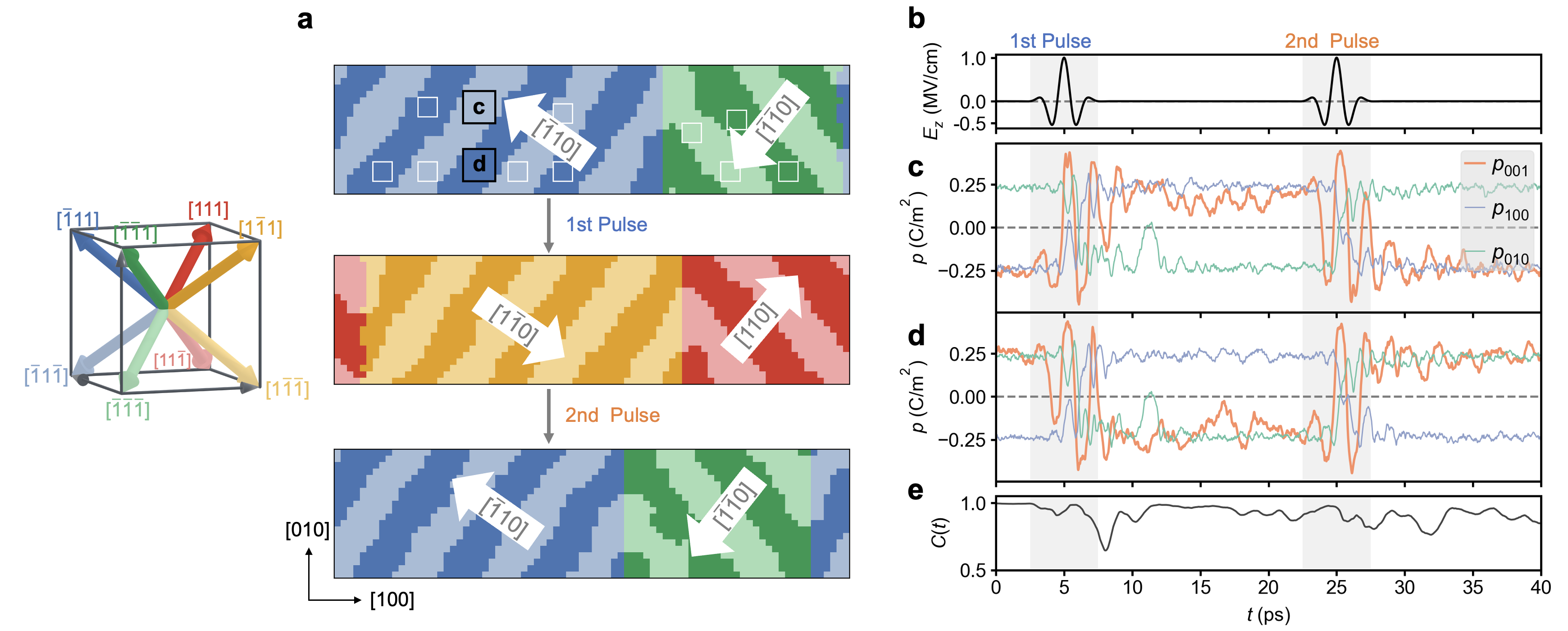}
\caption{
\textbf{Topology reversal of a double-$\mathbf q$ dipole wave induced by a single THz pulse.}
(a) Top views of the dipole-wave texture before excitation, after the first pulse, and after the second pulse. The two domain bands initially have in-plane polarization components along $[\bar{1}10]$ and $[\bar{1}\bar{1}0]$. The first out-of-plane THz pulse reverses these components to $[1\bar{1}0]$ and $[110]$, respectively. Most dipoles also reverse their out-of-plane components, corresponding to local 180$^\circ$ switching. The second pulse restores the original texture.
(b) THz pulse waveform. Both pulses have a peak field of $E_z=1$~MV/cm, a carrier frequency of 0.5~THz, and a Gaussian-envelope FWHM of 2~ps. Shaded regions indicate the temporal windows of the pulses.
(c,d) Time evolution of the local polarization $p$ in two representative regions marked by black boxes in the upper panel of (a). The first pulse reverses the local polarization from $[\bar{1}1\bar{1}]$ to $[1\bar{1}1]$ in (c) and from $[\bar{1}11]$ to $[1\bar{1}\bar{1}]$ in (d), despite their initially opposite out-of-plane components. The second pulse restores both regions to their initial polarization states.
(e) Time-dependent dipole--dipole correlation $C(t)$ calculated from the regions marked by white boxes in the upper panel of (a). The correlation remains substantial throughout the switching process, supporting the collective nature of the reversal.
The simulated nanocapacitor has an in-plane supercell size of $80\times20$ and contains 113,600 atoms in total.
}
\label{fig:pulse}
\end{figure*}

We now use the double-$\mathbf q$ dipole wave shown in Fig.~\ref{fig:md}d as an example to demonstrate that a single THz pulse can reverse polar topology. This texture consists of four degenerate $R$-like variants, $[\bar{1}11]$, $[\bar{1}1\bar{1}]$, $[\bar{1}\bar{1}1]$, and $[\bar{1}\bar{1}\bar{1}]$, separated by 71$^\circ$ domain walls. These variants form two domain bands distinguished by their in-plane polarization directions, $[\bar{1}10]$ and $[\bar{1}\bar{1}0]$ (Fig.~\ref{fig:pulse}a, upper panel). Within each band, variants with opposite out-of-plane polarization alternate periodically (\textit{e.g.}, $[\bar{1}11]$ and $[\bar{1}1\bar{1}]$). Fourier analysis confirms the double-$\mathbf q$ character, with peaks at $\mathbf q=(1/13,\pm1/20)$ (Fig.~S7). We focus on this texture because its out-of-plane polarization can couple directly to an electric field applied across the capacitor electrodes, while the associated $R$-like state can rotate relatively easily within the soft energy landscape.

The out-of-plane Gaussian-enveloped single-cycle THz pulse has the form 
$E_z(t)=E_0\cos\left[2\pi f(t-t_0)\right]
\exp\left[-4\ln 2 (t-t_0)^2/\tau^2 \right]$,
where $E_0$ is the peak field, $f$ is the carrier frequency, $t_0$ is the pulse center, and $\tau$ is the full width at half maximum (FWHM) of the Gaussian envelope. Unless otherwise noted, we use $f=0.5$~THz, $E_0=1$~MV/cm, and $\tau=2$~ps, as plotted in Fig.~\ref{fig:pulse}b. 
Figure~\ref{fig:pulse}a shows MD snapshots of the ultrafast switching process. Although the initial double-$\mathbf q$ dipole wave has nearly zero net out-of-plane polarization, a single pulse reverses most local dipoles by 180$^\circ$ within $\approx 2$~ps while preserving the double-$\mathbf q$ topology. To quantify this process, we track the averaged local polarization $p$ in two representative regions, $c$ and $d$, marked in Fig.~\ref{fig:pulse}a. In region $c$, the polarization switches from $[\bar{1}1\bar{1}]$ to $[1\bar{1}1]$ (Fig.~\ref{fig:pulse}c). Interestingly, region $d$, whose initial $[\bar{1}11]$ polarization has an out-of-plane component aligned with the positive peak field, also undergoes a complete reversal to $[1\bar{1}\bar{1}]$ (Fig.~\ref{fig:pulse}d). This response cannot be explained by a quasistatic thermodynamic argument, since a positive field cannot flip a positive polarization component to a negative one.
Remarkably, a second identical THz pulse restores the initial dipole-wave state (Figs.~\ref{fig:pulse}a,c--d). Thus, identical THz pulses can reversibly toggle the system between two topologically equivalent double-$\mathbf q$ variants, demonstrating ultrafast polar-topology reversal. It is noted that this switching is robust against MD simulation parameters, including ensemble choice and supercell size (see Figs.~S8 and S9).

We further quantify the switching pathway using the time-dependent dipole-dipole correlation $C(t)$ between selected regions marked by white boxes in the upper panel of Fig.~\ref{fig:pulse}a (see Supplemental Sect.~VIII for details). This correlation measures the degree to which dipoles in different regions rotate coherently during switching, with larger values indicating more synchronized motion. As shown in Fig.~\ref{fig:pulse}e, $C(t)$ remains above $\approx 0.5$ throughout the process, indicating that the selected regions switch nearly simultaneously. 
Unlike conventional polarization reversal, which typically proceeds through nucleation of field-aligned domains and subsequent domain-wall motion~\cite{Liu16p360,Yang25p021042,khachaturyan24p024403}, the switching here is spatially homogeneous. The required field is also modest: a peak field of 1~MV/cm corresponds to only $0.4$~V across a 4-nm film. We attribute this ultrafast, low-voltage response to the shallow rotational barriers of the $R$-like state.

\begin{figure*}
\centering
\includegraphics[width=1\textwidth]{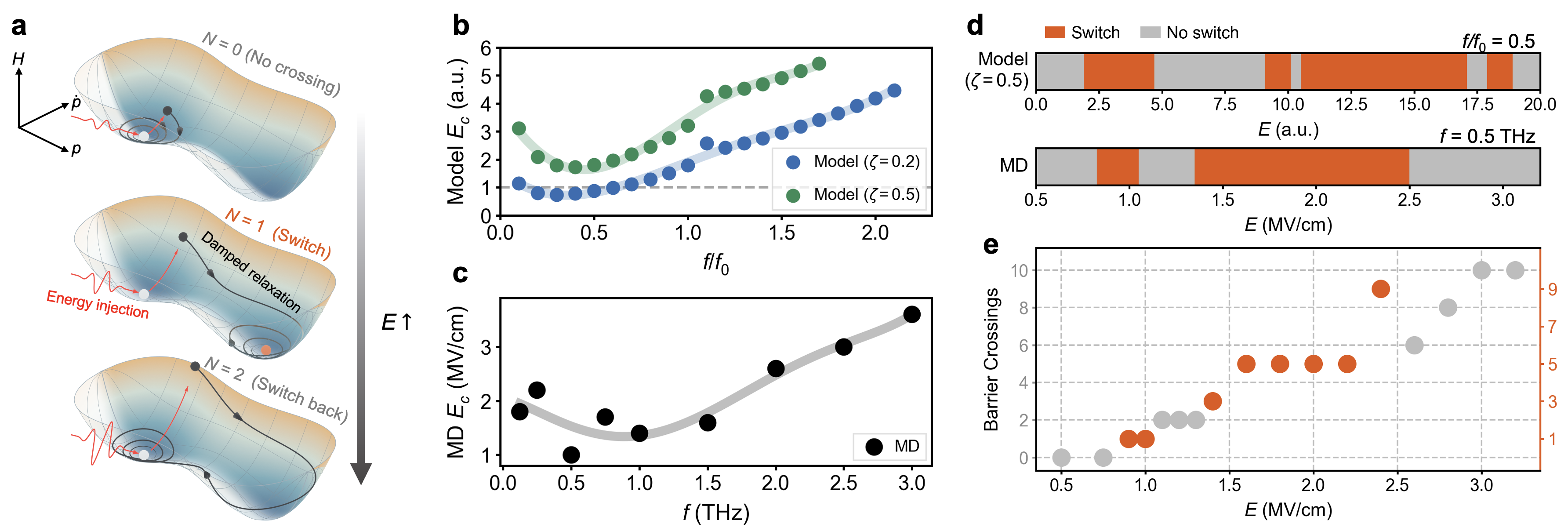}
\caption{
\textbf{Damping-assisted THz switching governed by inertial barrier-crossing dynamics.}
(a) Schematic phase-space illustration of THz-driven barrier crossing. Switching occurs when the trajectory crosses the barrier once and is trapped in the opposite well by damping; no switching occurs if the trajectory does not cross or crosses back before being trapped.
(b) Critical electric field $E_{\mathrm c}$ for switching predicted by the dipole model. The threshold shows a nonmonotonic, U-shaped frequency dependence. It increases with damping and reaches its minimum away from the nominal resonance at $f/f_0=1$.
(c) Critical switching fields from MD simulations of the double-$\mathbf q$ dipole wave, showing the same U-shaped frequency dependence.
(d) Switching and nonswitching windows obtained by scanning the field amplitude at fixed frequency for the dipole model (upper) and MD simulations (lower). Switching occurs only within selected amplitude windows.
(e) Number of barrier crossings as a function of field amplitude from MD simulations. The final state is set by the parity of the crossing number: an odd number gives switching, whereas an even number gives no switching after damping traps the polarization.
}
\label{fig:model}
\end{figure*}

% Fig 4
The out-of-plane THz field does not favor either the initial or reversed texture, because both have nearly zero net out-of-plane polarization. The reversal is therefore not driven by stabilization of a lower-energy state. We find that the switching occurs through a damping-assisted inertial mechanism, as sketched in Fig.~\ref{fig:model}a.  Specifically, a short THz pulse drives the polarization away from its initial energy minimum and can provide enough kinetic energy for it to cross a rotational barrier. After the pulse, damping removes the excess energy and traps the polarization in a nearby polar basin. If the polarization crosses the barrier once and is damped into the opposite basin, switching occurs (Fig.~\ref{fig:model}a, middle). If it fails to cross the barrier (Fig.~\ref{fig:model}a, top), or crosses back before being damped (Fig.~\ref{fig:model}a, bottom), the system returns to its original state. Therefore, successful reversal requires both sufficient energy injection and proper damping.
This mechanism thus has two main features. First, the critical field for polarization reversal depends on the THz frequency and has an optimal frequency. Second, even at the optimal frequency, a field above the critical value does not always guarantee reversal, because energy injection must be matched by dissipation. We demonstrate these two features using both a minimal dipole model and MD simulations.

In the dipole model, the polarization is treated as a fixed-magnitude vector rotating among symmetry-related [111]-type states, mimicking the local $R$-like state in the double-$\mathbf q$ dipole wave. Its orientation is parameterized by an angle $\theta$, with reversal defined by a sign change of $\theta$. The angular dynamics follow
$I\ddot{\theta}=-\partial U(\theta,t)/\partial\theta-\zeta\dot{\theta}$,
where $I$ is the effective moment of inertia, $U(\theta,t)$ includes the symmetric multiwell potential and coupling to the out-of-plane THz field, and $-\zeta\dot{\theta}$ represents dissipative damping. The THz pulse shape is the same as in MD simulations (see additional details in Supplemental Sect.~IX).

Numerical simulations based on the dipole model show that the same THz pulse can indeed reverse dipoles with opposite initial out-of-plane components, such as $[111]\rightarrow[\bar{1}\bar{1}\bar{1}]$ and $[11\bar{1}]\rightarrow[\bar{1}\bar{1}1]$ (see Fig.~S10). We define the critical field $E_{\mathrm c}$ as the lowest peak field required for reversal and calculate it as a function of THz frequency. Here, $E_{\mathrm c}$ is normalized by the characteristic field associated with the polarization-rotation barrier, and the frequency is normalized by $f_0$, the small-amplitude oscillation frequency near the local energy minimum.
As shown in Fig.~\ref{fig:model}b, $E_{\mathrm c}$ exhibits a shallow U-shaped dependence on frequency, indicating an optimal frequency for switching. At low frequencies, the field changes too slowly and much of the injected energy is dissipated during the pulse. At high frequencies, the pulse is too short to transfer energy efficiently. Between these limits, energy transfer to the polarization coordinate is most effective, producing a minimum in $E_{\mathrm c}$. For $\zeta=0.2$, the minimum occurs at $f/f_0=0.3$ with $E_{\mathrm c}=0.74$. Increasing the damping to $\zeta=0.5$ shifts the minimum to $f/f_0=0.4$ and raises it to $E_{\mathrm c}=1.73$, reflecting greater dissipative energy loss. Because these optimal frequencies are well below $f_0$, the switching is not governed by resonance with a polar mode (Fig.~S11).
We further use MD simulations to determine the critical field for topology reversal of the double-$\mathbf q$ dipole wave. The MD results show the same overall trend (Fig.~\ref{fig:model}c), with a minimum threshold of $E_{\mathrm c}=0.9$~MV/cm at 0.5~THz. This agreement confirms the first key feature of THz-driven inertial switching, namely the existence of an optimal frequency.

Second, both the dipole model and MD simulations show alternating switching and nonswitching windows as the field amplitude increases (Fig.~\ref{fig:model}d), demonstrating that larger fields are not always more effective. For example, in the MD simulations at 0.5~THz, a peak field of 0.9~MV/cm reverses the topology, whereas a stronger 1.2~MV/cm pulse eventually drives the dipole wave back to its original state (Fig.~S12).
This behavior reflects the need to balance energy input and dissipation. If the pulse deposits too much energy and damping is insufficient, the polarization can cross the barrier again and switch back. We confirm this in MD by counting the barrier crossings of the in-plane components (Fig.~\ref{fig:model}e). The number of crossings generally increases with field amplitude, and the final state is determined by its parity: an odd number of crossings gives the reversed state, whereas an even number returns the system to the original variant.
This damping-assisted inertial switching is reminiscent of precessional switching in magnetic systems, where a short field or torque pulse drives large-angle magnetization motion and damping selects the final magnetic basin~\cite{back99p864,gerrits02p509,schumacher03p017201}. We identify an analogous mechanism for ferroelectric polarization, particularly suited for topology reversal because it can flip local polarizations without requiring a net thermodynamic driving force. 

% Conclusion
In summary, our results establish a simple metal/ferroelectric/metal nanocapacitor as a route to programmable polar topology. Electrode-controlled boundary conditions, combining incomplete screening and in-plane strain, drive PbTiO$_3$ from its conventional tetragonal state into orthorhombic- and rhombohedral-like states, which support a rich variety of polar textures, including vortex lattices, skyrmion lattices, and dipole waves. We further demonstrate nonvolatile polar topology reversal in the double-$\mathbf q$ dipole wave using a single THz pulse. The reversal relies on damping-assisted inertial switching, which collectively switches antiparallel local polarizations. These results place ultrafast polar topology control in an open-system dynamical setting, where the balance between injected energy and dissipation is critical. They also suggest a practical pathway toward reversible, THz-speed control of complex polar states in nanoscale ferroelectric devices.

\clearpage
\newpage
\begin{acknowledgments}
We acknowledge support from the Zhejiang Provincial Natural Science Foundation of China (LR25A040004) and National Natural Science Foundation of China (92370104).
Computational resources were provided by the Open Source Supercomputing Center of Scientific
Advanced Intelligence (S-A-I).
\end{acknowledgments}

%\section*{Data Availability}
%The data that support the findings of this article are openly available~\cite{}.

\bibliography{SL.bib}

\clearpage
\newpage

\clearpage
\newpage

\clearpage
\newpage

\clearpage
\newpage

\end{document}